\documentclass[preprint,eqsecnum,preprintnumbers,nofootinbib,byrevtex,prd,aps,showpacs,showkeys,groupedaddress,
floatfix]{revtex4}
\usepackage{bm}
\usepackage{graphics}
\usepackage{graphicx}
\usepackage{epsfig}
\usepackage{amssymb}
\usepackage{amsmath}

\newcommand\nn{\nonumber}
\newcommand\ba{\begin{eqnarray}}
\newcommand\ea{\end{eqnarray}}

\begin{document}
%\large
\title{ Bound State Solution Schr\"{o}dinger Equation for Extended Cornell Potential at Finite Temperature}
\author{A.~I.~Ahmadov$^{1,2}$~\footnote{E-mail: ahmadovazar@yahoo.com}}
\author{K.~H.~Abasova$^{1}$~\footnote{E-mail: kabasova1996@mail.ru}}
\author{M.~Sh.~Orucova$^{3}$~\footnote{E-mail: morucova@mail.ru}}
\affiliation{$^{1}$ Department of Theoretical  Physics, Baku State
University,\\ Z. Khalilov st. 23, AZ-1148, Baku, Azerbaijan}
\affiliation{$^{2}$ Institute for Physical Problems, Baku State
University,\\ Z. Khalilov st. 23, AZ-1148, Baku, Azerbaijan}
\affiliation{$^{3}$ Azerbaijan State  University of Economics,\\
Istiglaliyyat st. 22,  AZ1001 Baku, Azerbaijan}

\date{}
\begin{abstract}

In this paper, we study the finite temperature-dependent
Schr\"{o}dinger equation by using the Nikiforov-Uvarov method. We
consider the sum of the Cornell, inverse quadratic, and harmonic-
type potential as the potential part of the radial Schr\"{o}dinger
equation. Analytical expressions for the energy eigenvalues and the
radial wave function are presented. Application of the results for
the heavy quarkonia and $B_c$ meson masses are good agreement with
the current experimental data except for zero angular momentum
quantum numbers. Numerical results for the temperature dependence
indicates a different behaviour for different quantum numbers.
Temperature- dependent results are in agreement with some QCD sum
rule results from the ground states.
\end{abstract}


  \maketitle

\section{Introduction}

It is a well-known fact that the potential models in quantum
mechanics are very accurate in reproducing experimental data for
meson spectroscopy \cite{Appelquist75} at zero temperature. However,
one needs to consider spin-dependent potentials in the
Schr\"{o}dinger equation in order to describe the  relativistic
effects \,\,~\cite{Greiner,Bagrov}. There exist few potentials which
are highly important because of their exact solubility within the
Schr\"{o}dinger equation for which all spectra of radial  $n_r$ and
orbital $l$ quantum  states can be obtained analytically
\,\,~\cite{Greiner,Bagrov,Landau,Dong}. Except the exact solvable
potentials, others are solved by either approximation or numerical
methods.

Several potentials such as the exponential-type  including the
Hulth\'{e}n,  Manning-Rosen, Woods-Saxon, and  Eckart-type potential
are also currently being investigated by several researchers. Among
the particularly interesting potentials which play an important role
in the quark-anti-quark bound states include the so-called Cornell
potential and a mixture of it with  the harmonic oscillator
potential and Morse potential as  discussed in
\cite{Kuchin,Al-Jamel,Maksimenko,Ghalenovi,Vigo-Aguir}.

As an analytical method, the Nikiforov-Uvarov (NU) method is one of
the widely applicable methods for solving the Schr\"{o}dinger
equation. The quarkonia system in a hot and dense matter media is
studied in Ref.\cite{Ahmadov1}, where the authors studied the
quarkonium dissociation in anisotropic plasma in hot and dense media
by analytically solving the multidimensional Schr\"{o}dinger
equation via the NU  method for the real part of the potential. The
NU method was successfully applied  for solving  the radial
Schr\"{o}dinger equation in the presence of an external magnetic
field and the Aharonov-Bohm flux fields in \cite{Karayer1,Karayer2}.
The inverse square root potential, which is a long-range potential
and a combination of the Coulomb, linear, and harmonic potentials,
is often used to describe quarkonium states.

The study of the heavy quark resonances in a nonrelativistic regime
and the thermal environment shows the importance of the color
screening radius below which binding is impossible~\cite{Matsui}.
The theoretical investigation of this effect for the charmonium resonance
is investigated in ~\cite{Karsch}. Considering the finite temperature for the Cornell
potential within the D-dimensional Schr\"{o}dinger equation by using the
NU method is presented in \,\,~\cite{Aby-Shady}. However, the
behaviour of bound states of heavy quarks in a strongly interacting
medium close to the deconfinement temperature $T_c$ is  largely
uncertain such that various models predict the mass being constant,
increasing and decreasing with temperature increments. One of the
first applications of a nonrelativistic lattice QCD to the study of
heavy quarkonia at finite temperature is presented in
\cite{Fingberg}.

Temperature-dependent Schr\"{o}dinger equations  for different
potentials by different methods are studied in
~\cite{El-Naggar,Malik,Wu,Maireche}. In Ref.\cite{Ahmadov2} a modified
radial Schr\"{o}dinger equation for the sum of the Cornell and inverse
quadratic potentials at finite temperature is solved. In Refs.\cite{
Abu-Shady3,Abu-Shady4}, the radial and hyperradial Schr\"{o}dinger
equations are analytically solved using the NU and SUSYQM methods, in
which the heavy quarkonia potential is introduced at finite
temperature with the baryon chemical potential. For numerical
solutions, one may, for instance, have a look in
\,\,\cite{Alberico:2006vw}.

The Cornell potential is extensively used to describe the mass
spectrum of the heavy quark and antiquark systems at zero
temperature ~\cite{Eichten1,Eichten2}. It is the sum of linear and
Coulomb terms which are responsible for the confinement and
quark-anti-quark interaction at short distances, respectively.

The bound state solutions to the wave equations under the
quark-antiquark interaction potential such as the ordinary,
extended, and generalized Cornell potentials and combined
potentials such as the Cornell with other potentials have attracted
much research interest in atomic and high - energy physics within
ordinary and supersymmetric quantum mechanics methods as in
\cite{Rani,Abu-Shady1,Vega,Khokha,
Ezz-Alarab,Gamal,Abu-Shady2,Hassanabadi}.

Recently, Ikot et al.\cite{Ikot} reported the approximate solutions
of the Schr\"{o}dinger equation with the central generalized
Hulth\'en and Yukawa potential  within the framework of the
functional method. The obtained wave function and the energy levels
are used to study the Shannon entropy, the Renyi entropy, the Fisher
information, the Shannon Fisher complexity, the Shannon power, and the
Fisher-Shannon product in both position and momentum spaces for the
ground and first excited states.

The exact solutions of the  Schr\"{o}dinger equation for the new
anharmonic oscillator, double ring-shaped oscillator, and  quantum
system with a nonpolynomial oscillator potential related to the
isotonic oscillator also  widely studied in
Refs.\cite{Dong3,Chen3,Dong4}. The relativistic Levinson theorem  was
also studied in Ref. \cite{Dong5}, and  the authors  obtained the
modified relativistic Levinson theorem for noncritical cases.

It is still an open question if the appropriate potential which
describes the interaction between a quark and an antiquark can be found
more precisely. It would be interesting to test the following
potential for the arbitrary orbital  quantum number $l\neq 0$  at
finite temperature by using NU method
 \,\,~\cite{Nikiforov}:

\begin{align}
    V(r)\equiv A\cdot
r-\frac{B}{r}+\frac{C}{r^2}+D\cdot r^2\,, \label{V}
\end{align}

\noindent Here, $A$, $B$, $C$, and $D$ are constant potential
parameter, respectively.

The rest of the paper is organized as follows. Temperature-dependent
radial Schr\"{o}dinger equation for the sum of the Cornell, inverse
quadratic, and harmonic oscillator-type potentials is introduced in
Section~ \ref{2} and solved using the NU method in  Section~  \ref{3}  .
In Section~ \ref{4},  we apply the results to the mass spectrum of
heavy mesons at zero and nonzero temperature in  Sec.~  \ref{5},
respectively. Finally, we end up with some concluding remarks in
Sec.~ \ref{6}.

\section{Temperature dependent Radial Schr\"{o}dinger
equation }\label{2}

The Schr\"{o}dinger equation in spherical coordinates is given as
follows:
\begin{align}
\nabla^2 \psi+ \frac{2\mu}{\hbar^2}[E-V(r)]\psi (r,\theta ,\phi )=0.
\label{a3} \end{align}

For the case of the separation of the wave function to the radial
and angular parts, we can write the wave function as follows:
\begin{align}\psi (r,\theta ,\phi
)=R(r)Y_{l,m}(\theta,\phi ), \label{a4}\end{align}

Consideration of the radial part of the wave function and the
Cornell, inverse quadratic, and harmonic-type potentials for the
potential part of the Schr\"{o}dinger equation leads to the
following:
\begin{align}R^{\prime\prime}(r)+\frac{2}{r}R'(r)+ \frac{2\mu}{\hbar ^2} \biggl[E
-\frac{l(l+1)\hbar^2}{2\mu r^2}-V(r)\biggr]R(r)=0.\label{a5}\end{align}

This equation gets the following form if we do the replacement for
the radial part of the wave function $R(r)\equiv \chi(r)/r$ in
(\ref{a5}):
%
%\begin{equation}
\begin{align}\chi^{\prime\prime}(r) +\frac{2\mu }{\hbar ^2}\left[E-\frac{\hbar
^2}{2\mu}\frac{l(l+1)}{r^2}-V(r)\right]\chi(r)= 0\,.
\label{a7}\end{align}
%\end{equation}
%
where $\mu$ is the reduced mass of the quark-antiquark system which
defined in this form:
$\frac{1}{\mu}=\frac{1}{\mu_q}+\frac{1}{\mu_{\overline{q}}}$.

Following the same philosophy  for the Cornell potential in
\cite{Fingberg}, one can do the nonzero temperature modification to
the constant terms in Eq.(\ref{V}) and make the potential term
temperature dependent as follows:
\begin{align}
V(T,r)\equiv A(T,r)\cdot
r-\frac{B(T,r)}{r}+\frac{C(T,r)}{r^2}+D(T,r)\cdot r^2\,,
\end{align}
where
\begin{align}
A(T,r)\equiv\frac{A}{\mu_{D}(T)\cdot
r}\Big(1-\exp\big\{-\mu_{D}(T)r\big\}\Big),\,\,B(T,r)\equiv
B\exp\big\{-\mu_{D}(T)r\big\},\\\nonumber C(T,r)\equiv
C\exp\big\{-\mu_{D}(T)r\big\},\,\,D(T,r)\equiv\frac{D}{\mu_{D}(T)\cdot
r}\Big(1-\exp\big\{-\mu_{D}(T)r\big\}\Big)\,.
\end{align}
Here, $\mu_{D}(T)$ is the Debye screening mass, which vanishes at
$T\rightarrow 0$. It should be noted that, in this model, the
temperature dependence of a potential is contains in a Debye
screened mass. Next, using the approximation
$exp(-\mu_{D}(T)r)=\sum_{n=0}^{\infty}\frac{(-\mu_{D}(T)r)^n}{n!}$
up to second order, which  gives a good accuracy when
$\mu_{D}(T)r<<1$, we then obtain the following:
\begin{align}
A(T,r)\equiv\frac{A}{\mu_{D}(T)\cdot
r}\Big(1-\exp\big\{-\mu_{D}(T)r\big\}\Big)=A-\frac{A}{2}\mu_{D}(T)r,
\label{a70} \nn
\end{align}
\begin{align}B(T,r)\equiv
B\exp\big\{-\mu_{D}(T)r\big\}=B(1-\mu_{D}(T)r+\frac{1}{2}\mu_{D}^{2}(T)r^2),
\end{align}
\begin{align}
C(T,r)\equiv
C\exp\big\{-\mu_{D}(T)r\big\}=C(1-\mu_{D}(T)r+\frac{1}{2}\mu_{D}(T)r^2),  \nn
\end{align}
\begin{align}
D(T,r)\equiv\frac{D}{\mu_{D}(T)\cdot
r}\Big(1-\exp\big\{-\mu_{D}(T)r\big\}\Big)=D-\frac{D}{2}\mu_{D}(T)r^2.
\nn
\end{align}
Then, for $V(T,r)$,  we obtain:
\begin{align}
V(T,r)=B\mu_{D}(T)+\frac{1}{2}C\mu_{D}^2(T)+(A-\frac{1}{2}B\mu_{D}^{2}(T))r-(B+C\mu_{D}(T))\frac{1}{r}\nonumber\\
-\Big(\frac{1}{2}A\mu_{D}(T)-D\Big)r^2+\frac{C}{r^2}-\frac{D}{2}\mu_{D}(T)r^3.\label{a16}
\end{align}
As can be seen from the expressions  in(\ref{a70}) at $T=0$ zero
temperature, $A(T=0)=A$,\,\,$B(T=0)=B$,\,\, $C(T=0)=C$ and
$D(T=0)=D$.

By using the expressions in (\ref{a70}) we may rewrite  expression
(2.5) in a more compact way as follows:
\begin{align}V(T,r)=F+Gr-\frac{L}{r}-Mr^2+\frac{C}{r^2}-Nr^3,\label{a21}\end{align}
where we used the following substitutions:
\begin{align}
F\equiv B\mu_{D}(T)+\frac{1}{2}C\mu_{D}^2(T),\,\,\,G\equiv
A-\frac{1}{2}B\mu_{D}^{2}(T),\\\nonumber
L\equiv B+C\mu_{D}(T),\,\,\,M\equiv
\frac{1}{2}A\mu_{D}(T)-D,\,\,\,N\equiv\frac{D}{2}\mu_{D}(T).
\label{a19}
\end{align}

Considering all these in the radial Schr\"{o}dinger equation
(\ref{a7}), we get the following:
\ba
%\begin{align}
\chi^{\prime\prime}(r)+\frac{2\mu }{\hbar^2}\biggl[E-\frac{\hbar^2}{2\mu}
\frac{l(l+1)}{r^2}-F-Gr+\frac{L}{r}+Mr^2-\frac{C}{r^2}+Nr^3\biggr]\chi (r)= 0.
\label{TScro}
%\end{align}
\ea
Let us reduce the above equation to  the generalized hypergeometric
type ~\cite{Nikiforov}:
\ba
%\begin{align}
\chi^{\prime\prime}(s)+\frac{\tilde{\tau}}{\sigma}
\chi'(s)+\frac{\tilde{\sigma}}{\sigma^2} \chi(s)=0.
\label{a34}
%\end{align}
\ea
In order to do that, we do the replacement $r=1/x$ in
Eq.(\ref{TScro}) which leads to the following:
%
%\ba
%\begin{align}
\ba \chi^{\prime\prime}(x) + \frac{2x}{x^2}\chi'(x)
+\frac{2\mu}{\hbar ^2}\frac{1}{x^4}\Big[E-\frac{\hbar
^2}{2\mu}l(l+1)x^2-F-\frac{G}{x}+Lx+\frac{M}{x^2}-Cx^2+\frac{N}{x^3}\Big]\chi
(x)= 0. \label{a25}\ea
%\end{align}

For the solution in Eq.(\ref{a25}), we introduce the following
approximation scheme on the term $G/x$, $M/x^2$ and $N/x^3$. Let us
consider a characteristic radius $r_0$ of the quark and  antiquark
system; it is the minimum interval between  two quarks at which they
cannot collide with each other. This scheme is based on the
expansion of $G/x$, $M/x^2$, and $N/x^3$ in a power series around
$r_0$ or $\delta=1/r_0$ in the x-space, up to the second order. One
should note that the $G$, $M$, and $N$ dependent terms, saves the original
form of equation (\ref{a25}). This approach is similar to the
Pekeris approximation~\cite{Pekeris}, which causes a deformation of
the centrifugal potential. Hence, after this modified potential,
Eq.(\ref{a25}) can be solved by the NU method. This expansion is
done for the new variable $y=x-\delta$ where $\delta=1/r_0$ around
$y=0$ as follows:

\ba
%\begin{align}
\frac{G}{x}=\frac{G}{y+\delta}=\frac{G}{\delta}(1-\frac{y}{\delta}+\frac{y^2}{\delta^2})=
\frac{G}{\delta}(1-\frac{x-\delta}{\delta}+\frac{(x-\delta)^2}{\delta^2})=
G\left(\frac{3}{\delta}-\frac{3x}{\delta^2}+\frac{x^2}{\delta^3}\right),
\label{a27}
%\end{align}
\ea
\ba
%\begin{align}
\frac{M}{x^2}=\frac{M}{(y+\delta)^2}=\frac{M}{\delta^2}(1+\frac{y}{\delta})^{-2}=
M\left(\frac{6}{\delta^2}-\frac{8x}{\delta^3}+\frac{3x^2}{\delta^4}\right),
\label{a28}
%\end{align}
\ea
\ba
%\begin{align}
\frac{N}{x^3}=\frac{N}{(y+\delta)^3}=\frac{N}{\delta^3}(1+\frac{y}{\delta})^{-3}=
N\left(\frac{10}{\delta^3}-\frac{15x}{\delta^4}+\frac{6x^2}{\delta^5}\right).
\label{a28}
%\end{align}
\ea

We get the following equation by substituting
Eqs.(\ref{a27}-\ref{a28}) into Eq.(\ref{a25}):
\ba
%\begin{align}
\chi^{\prime\prime}(x) + \frac{2x}{x^2}\chi^\prime(x)+\frac{2\mu }{\hbar
^2}\frac{1}{x^4}\biggl[\Big(E-F-\frac{3G}{\delta}+\frac{6M}{\delta^2}+\frac{10N}{\delta^3}\Big)+\Big(\frac{3G}{\delta^2}+L-
\frac{8M}{\delta^3}-\frac{15N}{\delta^4}\Big)x  \nonumber \\
+\Big(-\frac{\hbar^2}{2\mu}l(l+1)-\frac{G}{\delta^3}+\frac{3M}{\delta^4}+\frac{6N}{\delta^5}-C\Big)x^2\biggr]\chi (x)= 0. \,\,\,\,
\label{a29}
%\end{align}
\ea

In  Eq.(\ref{a29}), we introduce new variables for making the
differential equation more compact:
\ba
%\begin{align}
\label{a32}
H \equiv -\frac{2\mu }{\hbar
^2}\Big(E-F-\frac{3G}{\delta}+\frac{6M}{\delta^2}+\frac{10N}{\delta^3}\Big),\,\,\,P\equiv\frac{2\mu
}{\hbar ^2}\Big(\frac{3G}{\delta^2}+L-
\frac{8M}{\delta^3}-\frac{15N}{\delta^4}\Big),\\\nonumber
Q\equiv\frac{2\mu }{\hbar
^2}\Big(-\frac{\hbar^2}{2\mu}l(l+1)-\frac{G}{\delta^3}+\frac{3M}{\delta^4}+\frac{6N}{\delta^5}-C\Big).
%\end{align}
\ea

Finally, Eq.(\ref{a29}) gets the following more compact form:
\ba
%\begin{align}
\chi^{\prime\prime}(x)
+\frac{2x}{x^2}\chi^{\prime}(x)+\frac{1}{x^4}\left[-H+Px+Qx^2\right]\chi(x)= 0.
\label{a33}
%\end{align}
\ea
\section{NU Method Application}\label{3}

In this section, we will apply NU method for defining the energy
eigenvalues. A comparison of Eq.(\ref{a33}) and Eq.(\ref{a34}) leads
us to the following  redefinitions:
\ba
%\begin{align}
\tilde{\tau}
(x)=2x,\,\, \sigma(x)=x^2,\,\,\, \tilde{\sigma}(x)=(-H+Px+Qx^2).
\label{a35}
%\end{align}
\ea

\noindent Consider the following factorization:
\ba
%\begin{align}
\chi (x)=\phi (x)y(x)\,,
\label{a36}
%\end{align}
\ea
For the appropriate function $\phi (x)$, Eq.(\ref{a33}) takes the
form of the well-known hypergeometric-type equation. The appropriate
$\phi (x)$ function has to satisfy the following condition:
\ba
%\begin{align}
\frac {\phi^{\prime} (x)}{\phi (x)}=\frac {\pi (x)}{\sigma (x)},
\label{a37}
%\end{align}
\ea
\noindent where function $\pi (x)$ is the maximum degree of a polynomial with one variable and is defined as follows:
\ba
%\begin{align}
\pi(x)= \frac{{ \sigma^{\prime} -\tilde{\tau }}}{2} \pm \sqrt
{\left(\frac{{ \sigma^{\prime} -\tilde{\tau} }}{2}\right)^2 -\tilde{\sigma}
+k\sigma }=\frac{2x -2x}{2}\pm \sqrt{H-Px-Qx^2+kx^2}=\nonumber\\ \pm \sqrt{(k-Q)x^2-Px+H}.
\label{a38}
%\end{align}
\ea
Finally, we get the hypergeometric-type equation:
\ba
%\begin{align}
\sigma (x)y^{\prime\prime} (x) + \tau (x) y^{\prime} (x) +\overline{\lambda}\, y(x)=0\,,
%\end{align}
\ea
where $\overline{\lambda}$ and $ \tau (x)$ read
\ba
%\begin{align}
\overline{\lambda} =k+\pi ^{\prime}(x),\,\,\,\,\, \tau(x)=\tilde{\tau}(x) +2\pi(x)\,.
\label{a39}
%\end{align}
\ea
The constant parameter  $k$  can be defined by utilizing the
condition that the expression under the square root has a double
zero, i.e., its discriminant is equal to zero. Hence, we obtain the
following:
\ba
%\begin{align}
k =\frac{1}{4H}(P^2+4HQ)\,.
\label{a40}
%\end{align}
\ea
Now, substituting Eq.(\ref{a40}) into Eq.(\ref{a38}) leads us to the
following expression for $\pi(x)$:
\ba
%\begin{align}
\pi(x)= -\frac{1}{2\sqrt H}(Px-2H)\,.
\label{a41}
%\end{align}
\ea
According to the NU method, out of the two possible forms of the
polynomial $\pi(x)$, we select the one for which the function $\tau
(x)$  has the negative derivative. Another form is not suitable for
physical reasons. Therefore, the suitable functions for $\pi(x)$ and
$\tau(x)$ have the following forms:
\ba
%\begin{align}
\pi(x)= -\frac{1}{2\sqrt H}(Px-2H),\,\,\,\text{and}\,\,\, \tau(x)= 2x- \frac{Px}{\sqrt H}+2\sqrt H\,,
\label{a42}
%\end{align}
\ea
and their derivatives are as follows:
\ba
%\begin{align}
\pi^{\prime}(x)=-\frac{P}{2\sqrt H},\,\, \text{and}\,\,\,\tau^\prime(x)=2- \frac{P}{\sqrt H}\,.
\label{a43}
%\end{align}
\ea
We can define the constant $\overline{\lambda}$ from  Eq.(\ref{a39})
which reads as follows:
\ba
%\begin{align}
\overline{\lambda}=\frac{P^2}{4H}+Q-\frac{P}{2\sqrt H}\,.
\label{a48}
%\end{align}
\ea
Given a non negative integer $n_r$, the hypergeometric-type equation
has a unique polynomial solution of degree $n$ if
\ba
%\begin{align}
\overline{\lambda}=\overline{\lambda}_n=-n\tau^{\prime}-\frac{n(n-1)}{2}\sigma ^{\prime\prime},
\,\,\,\text{for}\,\, n=0,1,2...
\label{a49}
%\end{align}
\ea
with the condition $\overline{\lambda}_m\neq \overline{\lambda}_n$ for $m=0,1,2,...,n-1$. Furthermore, it follows that
\ba
%\begin{align}
\overline{\lambda} _{n_{r}}=-n_r(2-\frac{P}{\sqrt
H})-n_r(n_r-1)=-2n_r+\frac{P}{\sqrt
H}n_r-n_{r}^2+n_r=\frac{\sqrt{P}}{\sqrt H}n_r-n_r(n_r+1)\,,\label{a50}
%\end{align}
\ea
\ba
%\begin{align}
\frac{P^2}{4H}+Q-\frac{P}{2\sqrt H}=\frac{P}{\sqrt
H}n_r-n_r(n_r+1)\,.
\label{a51}
%\end{align}
\ea
We can solve Eq.(\ref{a51}) explicitly for $H$ and get the
following:
\ba
%\begin{align}
\sqrt H=\frac{P}{(1+2n)\pm\sqrt{1-4Q}}\,.
\label{a52}
%\end{align}
\ea
Substituting Eq.(\ref{a52}) into Eq.(\ref{a32}) we obtain the
following:
\ba
%\begin{align}
\sqrt {-\frac{2\mu }{\hbar
^2}\Big(E-F-\frac{3G}{\delta}+\frac{6M}{\delta^2}+\frac{10N}{\delta^3}\Big)}=
\frac{P}{(1+2n)\pm\sqrt{1-4Q}}\,.
\label{a53}
%\end{align}
\ea
From this equation, we can get the energy spectrum as follows:
%
%\ba
\begin{align}
E= F+\frac{3G}{\delta}-\frac{6M}{\delta^2}-\frac{10N}{\delta^3}- \\
\nn -\frac{\hbar ^2}{2\mu}\left[\frac{\frac{2\mu }{\hbar
^2}\Big(\frac{3G}{\delta^2}+L-
\frac{8M}{\delta^3}-\frac{15N}{\delta^4}\Big)}{(1+2n_r)+\sqrt{1+4l(l+1)+\frac{8\mu
}{\hbar ^2}\frac{G}{\delta^3}-\frac{24\mu }{\hbar
^2}\frac{M}{\delta^4}-\frac{48\mu }{\hbar
^2}\frac{N}{\delta^5}+\frac{8\mu }{\hbar ^2}C}}\right]^2.\label{a54}
\end{align}
%\ea
%
We would like to note that the $N$-dimensional radial
Schr\"{o}dinger equation for the same potential is solved in
~\cite{Ezz-Alarab}, which should be the same as our result for $N=3$
at zero temperature. Similar work with the Wentzel-Kramers-Brillouin
approximation method has been studied at $T=0$ temperature
in~\cite{Omugbe}. At $T=0$, zero temperature limit of the Eq.(3.17)
read as follows:
\begin{align}E=
\frac{3A}{\delta}+\frac{6D}{\delta^2}-\frac{\hbar ^2}{2\mu
}\left[\frac{\frac{2\mu }{\hbar ^2}\Big(\frac{3A}{\delta^2}+B+
\frac{8D}{\delta^3}\Big)}{(1+2n_r)+ \sqrt{1+4l(l+1)+\frac{8\mu
}{\hbar ^2}\frac{A}{\delta^3}+\frac{24\mu }{\hbar
^2}\frac{D}{\delta^4}+\frac{8\mu }{\hbar ^2}C}}\right]^2.
\end{align}
If we take  $C=0$ in Eq.(3.18), we  then obtain the following:
\begin{align}
E=\frac{3A}{\delta}+\frac{6D}{\delta^2}-\frac{\hbar ^2}{2\mu
}\left[\frac{\frac{2\mu }{\hbar ^2}\Big(\frac{3A}{\delta^2}+B+
\frac{8D}{\delta^3}\Big)}{(1+2n_r)+ \sqrt{1+4l(l+1)+\frac{8\mu
}{\hbar ^2}\frac{A}{\delta^3}+\frac{24\mu }{\hbar
^2}\frac{D}{\delta^4}}}\right]^2.
\end{align}
If we take $D=0$ in Eq.(3.18), we get \cite{Ahmadov2}
\begin{align}
E=\frac{3A}{\delta}-\frac{\hbar ^2}{2\mu
}\left[\frac{\frac{2\mu }{\hbar
^2}\Big(\frac{3A}{\delta^2}+B
\Big)}{(1+2n_r)+ \sqrt{1+4l(l+1)+\frac{8\mu }{\hbar
^2}\frac{A}{\delta^3}+\frac{8\mu }{\hbar
^2}C}}\right]^2.
\end{align}
If we take $C=0$ and $D=0$ in Eq.(3.18) we obtain the same result as
follows\cite{Kuchin}:
\begin{align}
E=\frac{3A}{\delta}-\frac{\hbar ^2}{2\mu }\left[\frac{\frac{2\mu
}{\hbar ^2}\Big(\frac{3A}{\delta^2}+B)}{(1+2n_r)+
\sqrt{1+4l(l+1)+\frac{8\mu}{\hbar^2}\frac{A}{\delta^3}}}\right]^2.
\end{align}

We can also find the radial eigenfunctions by applying the NU
method. The relevant $\pi(s)$  function must satisfy the following
condition:
\begin{align}\frac { \phi^{\prime} (x)}{\phi (x)}=\frac {\pi (x)}{\sigma
(x)}=\frac{\pm(\frac{Px}{2\sqrt H}-\sqrt
H)}{x^2}=\pm\left(\frac{P}{2\sqrt H x}-\frac{\sqrt H}{x^2}\right),
\label{a57}
\end{align}
It is not a complicated task to find the following result, after
substituting $\pi(x)$ and $\sigma(x)$ into Eq.(\ref{a57}) and
solving a first-order differential equation:
\begin{align}\phi (x)=x^{\frac{P}{2\sqrt H}}e^{-\frac{\sqrt
H}{x}}\,.
\label{a58}
\end{align}
Furthermore, the other part of the wave function $y_{n}(x)$ is the
hypergeometric-type function whose polynomial solutions are given by
Rodrigues relation:
\begin{align}y_{n}(x) = \frac {C_{n}}{\rho (x)} \frac{{d^{n} }}{{dx^{n}
}}\left[ \sigma ^{n}(x)\rho (x) \right],
\label{a59}
\end{align}
\noindent where $C_n$ is a normalizing constant and $\rho (x)$ is
the weight function which is the solution of the Pearson
differential equation. The Pearson differential equation and
$\rho(x)$ for our problem is given as follows:
\begin{align}
(\sigma \rho)^{\prime} =\tau \rho.
\label{a60}
\end{align}
Therefore, we use  equation(\ref{a60}) to find the  second part of
the wave function from the definition of weight function:
\begin{align}
\rho(x) =x^{-P/\sqrt H} e^{{-2\sqrt H}/x }.
\label{a61}
\end{align}
Considering both parts of the wave function $\phi (x)$ and
$y_{n_{r}}(x)$ within Eq.(\ref{a36}), we obtain the following:
\begin{align}
\chi _{n_{r}}(x)=C_{n_{r}l}\cdot\,x^{\frac{P}{2\sqrt
H}}\cdot\,e^{\frac{\sqrt H}{x}}
\cdot\,\frac{d^{n}}{dx^{n}}\left[x^{2n-\frac{P}{\sqrt H}}e^{-2\sqrt H/x}
\right].
\label{a62}
\end{align}
As the last step, we do the replacement $x=1/r$, and using
$\chi(r)=rR(r)$ in Eq.(\ref{a62}) we get the following:
\begin{align}
\chi_{n_{r}}(r)=C_{n_{r}l}\cdot\,r^{-\frac{P}{2\sqrt H}}e^{\sqrt H\cdot
r}\left(-r^2\frac{d}{dr}\right)^{n}\left[r^{-2n+\frac{P}{\sqrt
H}}e^{-2\sqrt H\cdot r} \right].
\label{a63}
\end{align}
The final form of the radial wave function $R(r)$ reads:
\begin{align}
R(r)=C_{n_{r}l}\cdot\,r^{-1-\frac{P}{2\sqrt H}}e^{\sqrt H\cdot r}
\left(-r^2\frac{d}{dr}\right)^{n}\left[r^{-2n+\frac{P}{\sqrt
H}}e^{-2\sqrt H\cdot r} \right].
\label{a64}
\end{align}
\section{Mass spectrum of the heavy quarkonium}\label{4}
We calculate the mass spectra of the heavy quarkonium system, for
example, charmonium  and bottomonium mesons that are the bound state
of  quarks and antiquarks. For this we apply the following relation:
\begin{align}
M=m_{q}+m_{\bar{q}}+E\,,
\label{a65}
\end{align}
\noindent where $m$ is  the bare mass of a heavy quark.
Using expression (\ref{a54}) for the energy spectrum in (\ref{a65})
we get the following equation for  heavy quarkonia mass at  finite
temperature :
\begin{align}\nonumber
M=m_{q}+m_{\overline{q}}+ F+\frac{3G}{\delta}-\frac{6M}{\delta^2}-\frac{10N}{\delta^3}-\\
-\frac{\hbar
^2}{2\mu }\left[\frac{\frac{2\mu }{\hbar
^2}\Big(\frac{3G}{\delta^2}+L-
\frac{8M}{\delta^3}-\frac{15N}{\delta^4}\Big)}{(1+2n)+\sqrt{1+4l(l+1)+\frac{8\mu
}{\hbar ^2}\frac{G}{\delta^3}-\frac{24\mu}{\hbar
^2}\frac{M}{\delta^4}-\frac{48\mu}{\hbar
^2}\frac{N}{\delta^5}+\frac{8\mu}{\hbar^2}C}}\right]^2
\label{a66}.
\end{align}
Depending on the system which we want to study, we may consider that
$m_q$ and $m_{\overline{q}}$ are the bare masses of quarks and
antiquarks correspondingly, $E$ is the energy of the system. By
replacing  $T=0$, we obtain the meson mass at zero temperature:
\begin{align}
M=m_{q}+m_{\overline{q}}+
\frac{3A}{\delta}+\frac{6D}{\delta^2}-\frac{\hbar ^2}{2\mu
}\left[\frac{\frac{2\mu }{\hbar ^2}\Big(\frac{3A}{\delta^2}+B+
\frac{8D}{\delta^3}\Big)}{(1+2n)+ \sqrt{1+4l(l+1)+\frac{8\mu }{\hbar
^2}\frac{A}{\delta^3}+\frac{24\mu}{\hbar
^2}\frac{D}{\delta^4}+\frac{8\mu }{\hbar ^2}C}}\right]^2
\label{a67}.
\end{align}
Numerical values for the charmonium mass
spectra is presented in Table \ref{Tab1}:

\begin{table}[h]
\begin{tabular}{|l|l|l|l|l|l|}
\hline
States & \begin{tabular}[c]{@{}l@{}}Present\\ Paper \end{tabular} & \begin{tabular}[c]{@{}l@{}}Experimental\\ Results  \cite{Zyla}\end{tabular} & States & \begin{tabular}[c]{@{}l@{}}Present\\ Paper \end{tabular} & \begin{tabular}[c]{@{}l@{}}Experimental \\ Results  \cite{Zyla}  \end{tabular} \\ \hline\hline
$J/\psi(1s)$     &   $3.098$   &      $3.097$      & $1p$     & $3.256$          &  $3.511$       \\ \hline
$\psi(2s)$     &  $3.687 $                                                            &   $3.686 $                                                             &  $2p$     &  $3.780$                                                             &       3.922                                                     \\ \hline
$3s- \psi(4040)$     & $4.042$                                                              &  $4.039$                      & $3p$     &   $4.100$                                                            &                           \\ \hline
$4s-\psi(4260)$     &    $4.272$                                                           &   $4.259$                                                              & $4p$     &  $4.310$                                                             &                                                             \\ \hline
$5s- \psi(4415)$    &     $4.429$                                                          & $4.421$                                                             & $5p$     &  $4.456$                                                             &                                                                 \\ \hline
$6s$     &       $4.540$                                                        &                                                            &  $1d$      &      $3.505$                                                         &   $3.774$
\\ \hline
\end{tabular}
        \vspace{0.5cm}
\caption{Mass spectra of charmonium resonances in
$\text{GeV}$.}\label{Tab1}
\end{table}
In this table, we considered the bare mass of the charm quark as
$m_c=1.209\,\,\text{GeV} $, and the constant terms fitted with the
experimental data via Eq.(\ref{a67}) as $A=0.2\,\,\text{GeV}^2$,
$B=1.244$, $C=2.9\times 10^{-3}$, $D=1.4\times 10^{-5}$, and
$\delta=0.231\,\,\text{GeV}$. If we apply formula (\ref{a67}) to the
bottomonium case with the bare mass $m_b=4.823\,\, \text{GeV}$, and
the experiment fitted constants use Eq.(\ref{a67}) as $A=0.2\,\,
\text{GeV}^2$, $B=1.569$, $C=2.0\times 10^{-3}$, $D=1.4\times
10^{-5}$, and $\delta=0.378\,\,\text{GeV}$, we obtain the following
results shown Table \ref{Tab2}.

\begin{table}[h]
\begin{tabular}{|l|l|l|l|l|l|}
\hline
States & \begin{tabular}[c]{@{}l@{}}Present\\ Paper\end{tabular} & \begin{tabular}[c]{@{}l@{}}Experimental\\ Results \cite{Zyla}\end{tabular} & States & \begin{tabular}[c]{@{}l@{}}Present\\ Paper\end{tabular} & \begin{tabular}[c]{@{}l@{}}Experimental \\ Results \cite{Zyla}\end{tabular} \\ \hline\hline
$\Upsilon(1s)$     &   $9.460$   &       $9.460$      & $1p$     & $9.619$          &  $9.899$       \\ \hline
$\Upsilon(2s)$      &  $10.023 $                                                            &   $10.023 $                                                             & $2p$     &  $10.114$                                                             &   $10.260$                                                              \\ \hline
$\Upsilon(3s)$   & $10.355$                                                              &  $10.355$                                                             & 3p     &   $10.411$                                                            &                                                                 \\ \hline
$\Upsilon(4s)$     &    $10.567$                                                           &  $10.579$                                                               & 4p     &  $10.604$                                                             &                                                                 \\ \hline
$7s-\Upsilon(10860)$      &     $10.887$                                                          &  $10.885$                                                           & 5p     &  $10.736$                                                             &                                                                 \\ \hline
$10s-\Upsilon(11020)$       &     $11.021$                                                          &   $11.020$                                                          &  $1d$      &  $9.863$                                                             &         $10.164$                                                        \\ \hline
\end{tabular}
    \vspace{0.5cm}
\caption{Mass spectra of bottomonium resonances in
$\text{GeV}$.}\label{Tab2}
\end{table}

Considering the $T=0$ and $D=0$ limit in Eq.(\ref{a66}) we get the
following expression which is obtained in \,\,~\cite{Ahmadov2}

\begin{align}
M=m_{q}+m_{\overline{q}}+ \frac{3A}{\delta}-\frac{\hbar ^2}{2\mu
}\left[\frac{\frac{2\mu }{\hbar ^2}\Big(\frac{3A}{\delta^2}+B
\Big)}{(1+2n)+ \sqrt{1+4l(l+1)+\frac{8\mu }{\hbar
^2}\frac{A}{\delta^3}+\frac{8\mu }{\hbar ^2}C}}\right]^2 \,.
\end{align}

$T=0$\,, $C=0$, and $D=0$ limits of equation (\ref{a66}) lead to the
following formula which  fully coincides with the  quarkonium mass
formula at zero temperature in \,\,~\cite{Kuchin}
\begin{align}
M=m_{q}+m_{\overline{q}}+ \frac{3A}{\delta}-\frac{\hbar ^2}{2\mu
}\left[\frac{\frac{2\mu }{\hbar
^2}\Big(\frac{3A}{\delta^2}\Big)}{(1+2n)+ \sqrt{1+4l(l+1)+\frac{8\mu
}{\hbar ^2}\frac{A}{\delta^3}}}\right]^2 .\end{align}

We may conclude that our current results as presented in
Tab.\ref{Tab1}  and Tab.\ref{Tab2} are in good agreement with
current available experimental data for all states of charmonium and
bottomonium resonances. The main reason why the results for the $p$
and $d$ states are not in good agreement with the experimental data
comes from the nonrelativistic calculation which we use throughout
our calculations. One needs to consider the spin-spin and
spin-orbital interactions terms within the potential.  Thus, the
reason is not related to the correct choice of the parameters or
making a better fit. It is impossible to consider the spin terms
within the Schr\"{o}dinger equation because of its nonrelativistic
nature. These terms should be considered within the relativistic
equations such as within the Klein-Fock-Gordon and Dirac equations.
We are planing to study it in the future since it is out of the
scope of the current paper.

In the Table \ref{Tab3}, we present the mass spectrum results for
the $B_c$ mesons with masses $m_c=1.209\,\, \text{GeV}$,
$m_b=4.823\,\, \text{GeV}$, and parameters   $A=0.147\,\,
\text{GeV}^2$, $B=1.204$, $C=2.8\times 10^{-3}$, $D=2.4\times
10^{-5}$,  and $\delta=0.379\,\,\text{GeV}$.

\begin{table}[h]
\begin{tabular}{|l|l|l|l|l|l|}
\hline States & \begin{tabular}[c]{@{}l@{}}Present\\ Paper\end{tabular} & \begin{tabular}[c]{@{}l@{}}Experimental\\Results \cite{Zyla}\end{tabular} & States &
\begin{tabular}[c]{@{}l@{}}Present\\ Paper\end{tabular}
\begin{tabular}[c]{@{}l@{}}
\end{tabular} \\ \hline\hline
$B_C^{+}~(1s)$  & $6.277$ & $6.275$  & $1p$   & $6.593$   \\ \hline
$B_C(2s)^{\pm}$  &  $6.763 $ &   $6.872$ & $2p$   &  $6.875$   \\ \hline
$3s$  & $6.945$ &  & 3p   &   $6.700$     \\ \hline
$4s$  &  $7.033$ &   & 4p     &  $7.061$   \\ \hline
$5s$  &  $7.081$ &    & 5p &  $7.098$    \\ \hline
$6s$  &  $7.111$ &  & $1d$ &  $6.831$     \\ \hline
\end{tabular}
    \vspace{0.5cm}
\caption{Mass spectra of $b\overline{c}$ resonances in
$\text{GeV}$.}\label{Tab3}
\end{table}

In order to compare our results with the different theoretical
works, we present Tables \ref{table1} and \ref{table2} for the mass
spectra of charmonium and bottomonium accordingly.

\begin{table}[h]
\begin{tabular}{|c|c|c|c|c|c|c|c|c|c|}\hline
 $\text{state} $& \text{Present paper} & $D=0$ \cite{Ahmadov2} & $C=0$ \cite{Kuchin} &\cite{Faustov}& \cite{Kumar}&\cite{Al-Jamel} &\text{Exp}. \cite{Zyla} \\ \hline

1s&3.098 &3.097&3.096 &3.068&3.078&3.096& 3.097 \\\hline
2s&3.687 &3.687&3.686 &3.697&3.581&3.686&3.686 \\  \hline
3s& 4.042 &4.041&4.040&4.144 &4.085 &3.984&4.039 \\\hline
4s&4.271 &4.271&4.269 & &4.589&4.150& 4.421\\ \hline
5s&4.428 &4.428&4.425&& && \\\hline
1p&3.256 &3.256&3.255&3.526 &3.415 &3.433& 3.511 \\ \hline
2p&3.780 &3.780&3.779&3.993&3.917 &3.910& 3.922\\ \hline
3p&4.100 &4.100 &&&&&  \\ \hline
4p&4.310 &4.310 & & & &&\\ \hline
5p&4.456 &4.456 & & & &&\\ \hline
1d&3.505 &3.505 &3.504 &3.829 &3.749 &3.767&3.774\\ \hline
\end{tabular}
\vspace{0.5cm}
\caption{Mass spectra of charmonium in GeV.}
\label{table1}
\end{table}
%\vspace*{5cm}
%\noindent
%\clearpage
%\newpage
\begin{table}[h]
\begin{tabular}{|c|c|c|c|c|c|c|c|c|c|}\hline
 $\text{state} $& \text{Present paper} & $D=0$ \cite{Ahmadov2} & $C=0$ \cite{Kuchin} &\cite{Faustov}& \cite{Kumar}&\cite{Al-Jamel} & \text{Exp}. \cite{Zyla} \\ \hline

  1s&9.460 &9.459&9.460&9.447&9.510&9.460& 9.460\\  \hline
  2s&10.023 &10.022 &10.023&10.012 &10.038 &10.023&10.023 \\ \hline
  3s&10.355 &10.354&10.355 &10.353&10.566 &10.280& 10.355\\ \hline
  4s&10.567 &10.566&10.567 &10.629&11.094&10.420& 10.579\\ \hline
  5s&10.710 &10.710 & && && \\ \hline
  1p& 9.619 &9.618&9.619&9.900&9.862 &9.840&9.899 \\ \hline
  2p&10.114 &10.113 &10.114&10.260 &10.390&10.160& 10.260 \\ \hline
  3p &10.411 &10.411 & & & &&\\ \hline
  4p&10.604 &10.604&&&&& \\  \hline
  5p&10.736 &10.736&& &&& \\ \hline
  1d&9.863 &10.257 &9.864 &10.155 &10.214 &10.140&10.164\\ \hline
\end{tabular}
\vspace{0.5cm}
\caption{Mass spectra of bottomonium in GeV.}
\label{table2}
\end{table}

\newpage\section{Temperature dependence of quarkonia mass}\label{5}

In this section, we present the numerical results for the
temperature dependence for some heavy meson mass spectra.

In Section IV, we have seen that the present potential model for
describing the heavy quarkonia ans $B_c$ mesons mass at zero
temperature is quite a good candidate. For studying the temperature
dependence, we will follow  \cite{Karsch}.

For calculation the mass  spectra at finite temperature, we use
the explicit form of Debye screening mass $\mu_{D}(T)$ according to
\cite{Boyd}:

\begin{align}
\mu_{D}(T)=\gamma\alpha_{s}(T)T.
\label{a68}
\end{align}

\noindent where $\gamma=14.652\pm0.337$.  In the numerical
calculations for the running coupling constant $\alpha_{s}(T)$, we
will adopt the following form at finite temperature:
\begin{align}
\alpha_{s}(T)=\frac{2\pi}{(11-\frac{2}{3}N_f)\log(\frac{T}{\Lambda})}.
\label{a69}
\end{align}
Here, we take the  critical temperature
$T_c=(169\pm16)\,\,\text{MeV}$ from lattice QCD  ~\cite{Bernard}
which leads to $\Lambda=\beta T_c=(17.6\pm3.2)\,\,\text{MeV}$.

In the numerical calculations, we will apply as $N_f = 3$ with two
light quarks of the same mass $u$ and $d$ and one heavier $s$.

\begin{figure}[h]
\includegraphics[width=30pc]{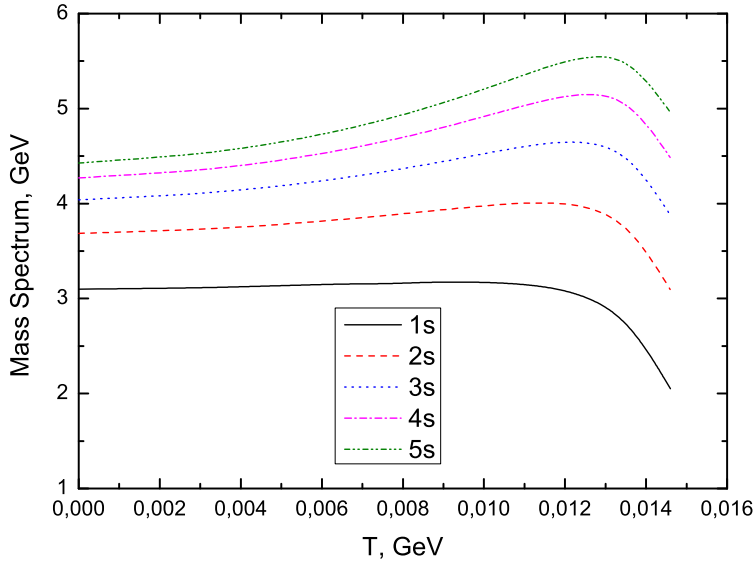}%\hspace{2pc}%
\caption{The mass spectrum of charmonium in the $1s$, $2s$, $3s$,
$4s$, and $5s$ states as a function  of the temperature $T$ with a
mass $m_c=1.209\,\, \text{GeV}$ and parameters of
$A=0.2\,\,\text{GeV}^2$, $B=1.244$, and
$\delta=0.231\,\,\text{GeV}$.}\label{fig1}
\end{figure}

\begin{figure}[h]
\includegraphics[width=30pc]{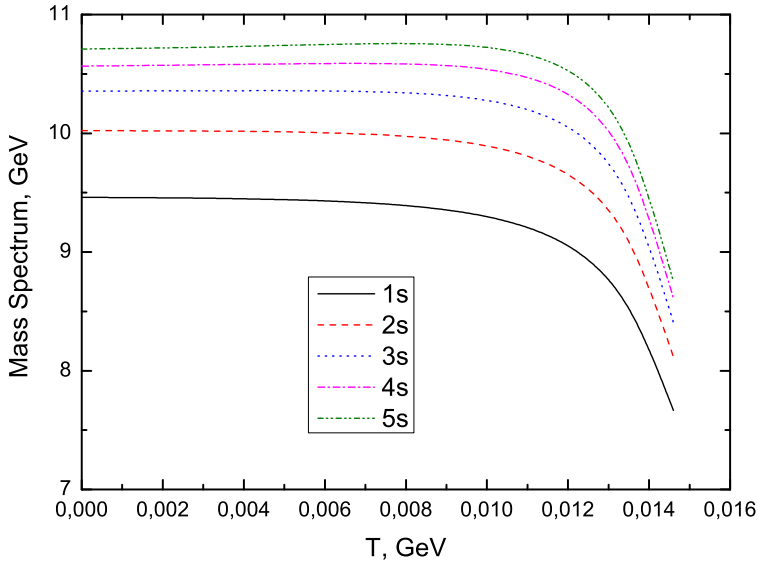}%\hspace{2pc}%
\caption{The mass spectrum of bottomonium  in the $1s$, $2s$, $3s$,
$4s$, and $5s$ states as a function  of the temperature $T$ with a
mass of $m_b=4.823\,\, \text{GeV}$ and parameters of $A=0.2\,\,
\text{GeV}^2$, $B=1.569$,and
$\delta=0.378\,\,\text{GeV}$.}\label{fig2}
\end{figure}

\begin{figure}[h]
\includegraphics[width=30pc]{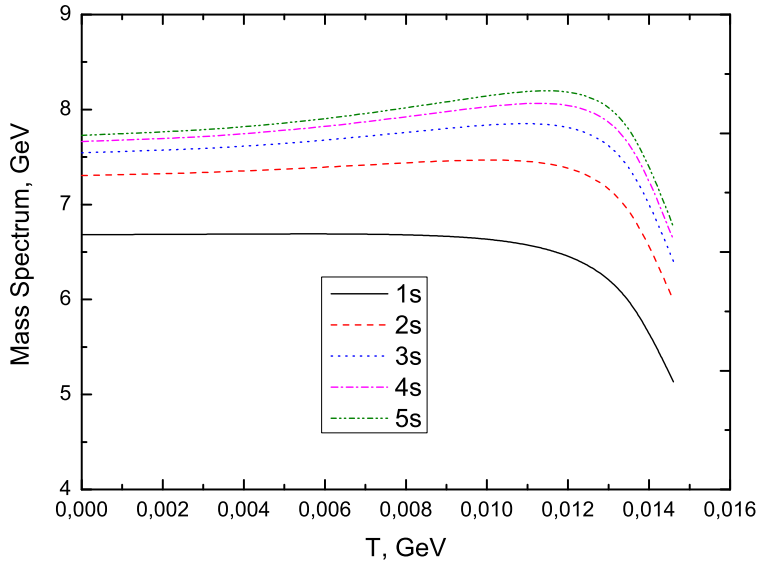}%\hspace{2pc}%
\caption{The mass spectrum of $b\overline{c}$ for $1s$, $2s$, $3s$,
$4s$, and $5s$ states as a function  of the temperature $T$ with
masses of $m_c=1.209\,\, \text{GeV}$ and $m_b=4.823\,\, \text{GeV}$,
and parameters  of $A=0.2\,\, \text{GeV}^2$, $B=1.407$, and
$\delta=0.324\,\,\text{GeV}$.}\label{fig3}
\end{figure}

\begin{figure}[h]
\includegraphics[width=30pc]{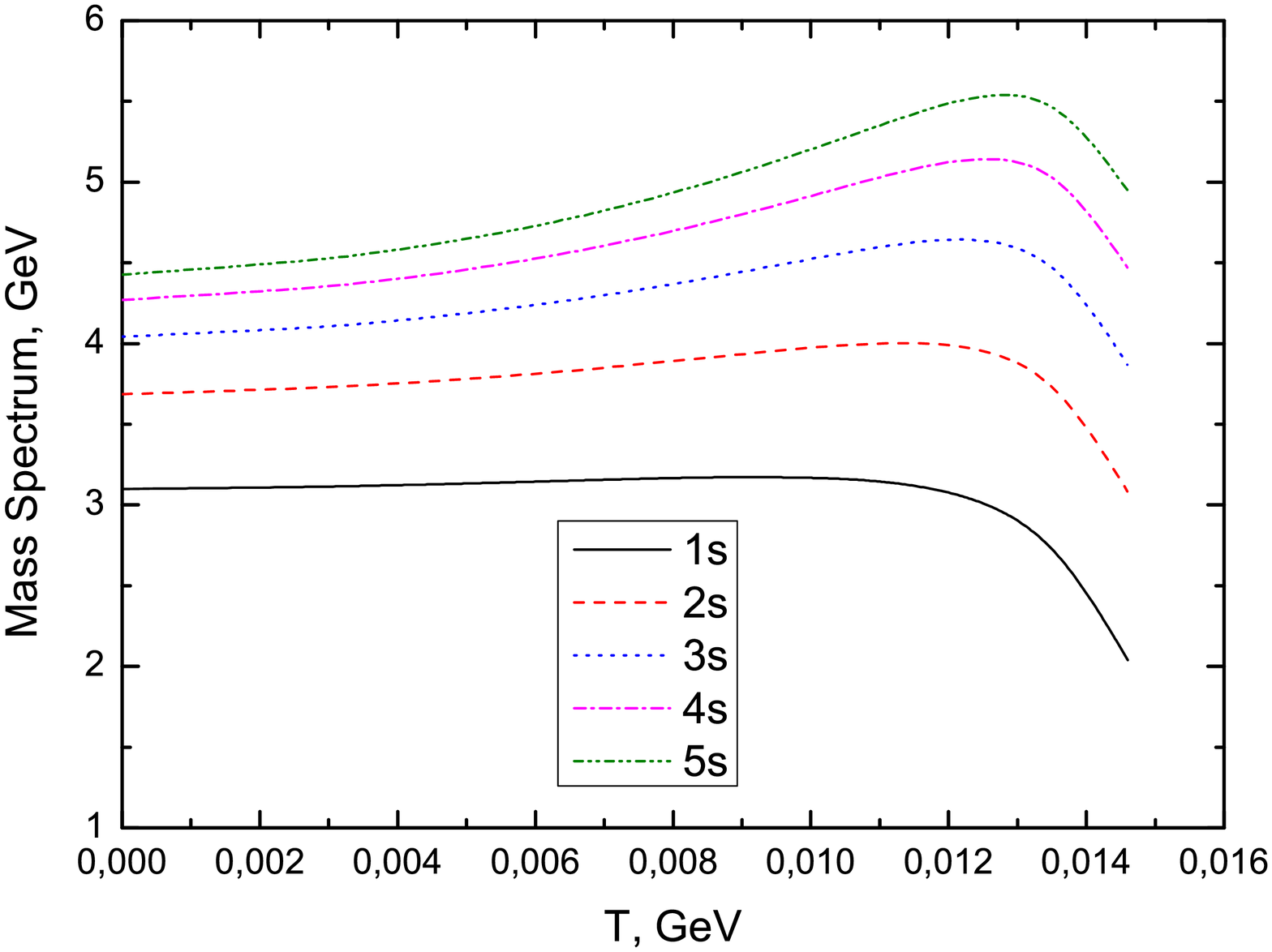}%\hspace{2pc}%
\caption{The mass spectrum of charmonium for $1s$, $2s$, $3s$, $4s$,
and $5s$ states as a function  of the temperature $T$ with a mass of
$m_c=1.209\,\, \text{GeV}$ and   parameters of $A=0.147\,\,
\text{GeV}^2$, $B=1.204$, $C=2.8\times 10^{-3}$, $D=2.4\times
10^{-5}$, and $\delta=0.379\,\,\text{GeV}$.}\label{fig4}
\end{figure}

\begin{figure}[h]
\includegraphics[width=30pc]{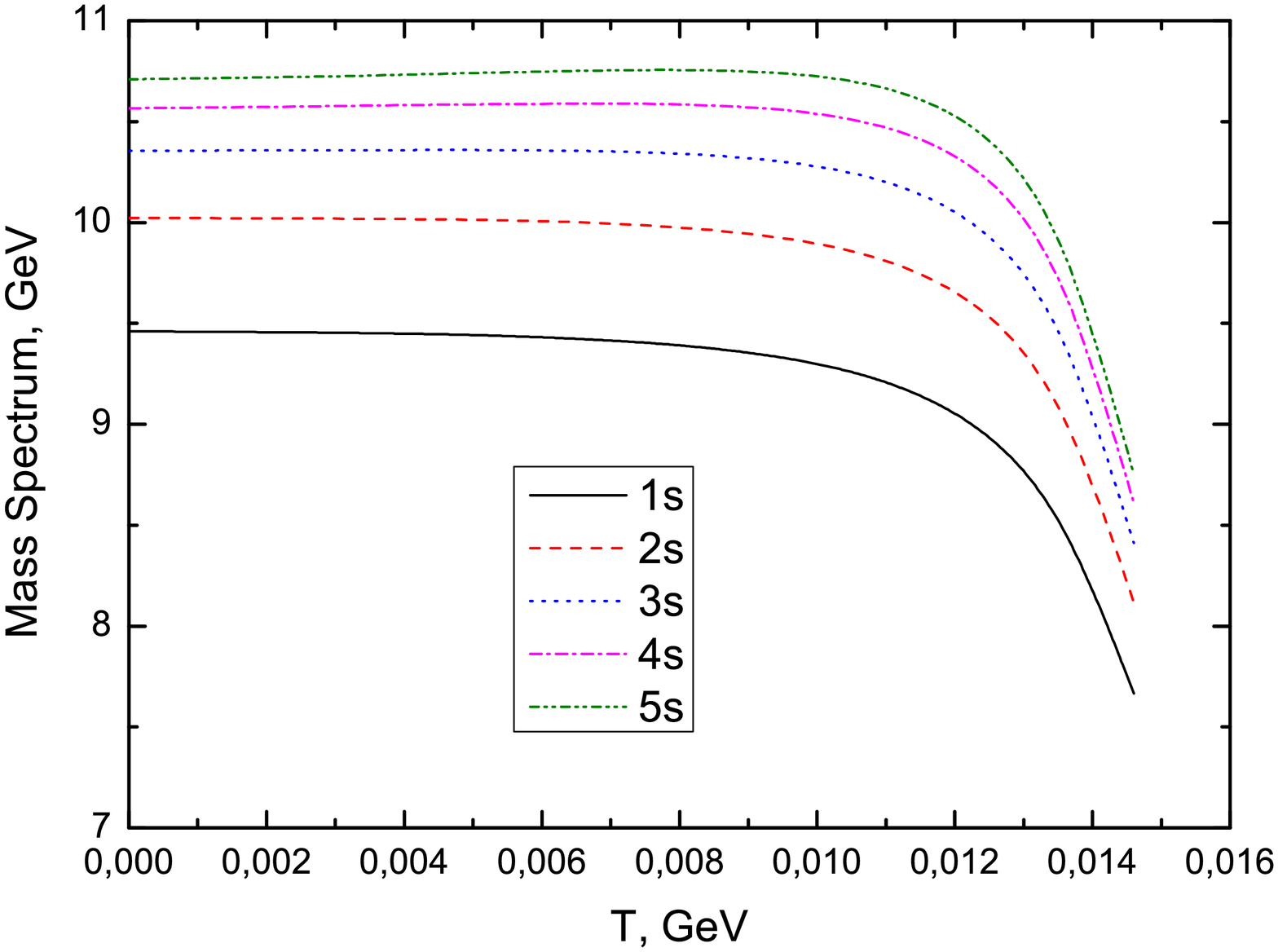}%\hspace{2pc}%
\caption{The mass spectrum of $bc$ for $1s$, $2s$, $3s$, $4s$, and
$5s$ states as a function  of the temperature $T$ with a mass of
$m_b=4.823\,\, \text{GeV}$ and parameters  $A=0.147\,\,
\text{GeV}^2$, $B=1.204$, $C=2.8\times 10^{-3}$, $D=2.4\times
10^{-5}$, and $\delta=0.379\,\,\text{GeV}$.}\label{fig5}
\end{figure}

\begin{figure}[h]
\includegraphics[width=30pc]{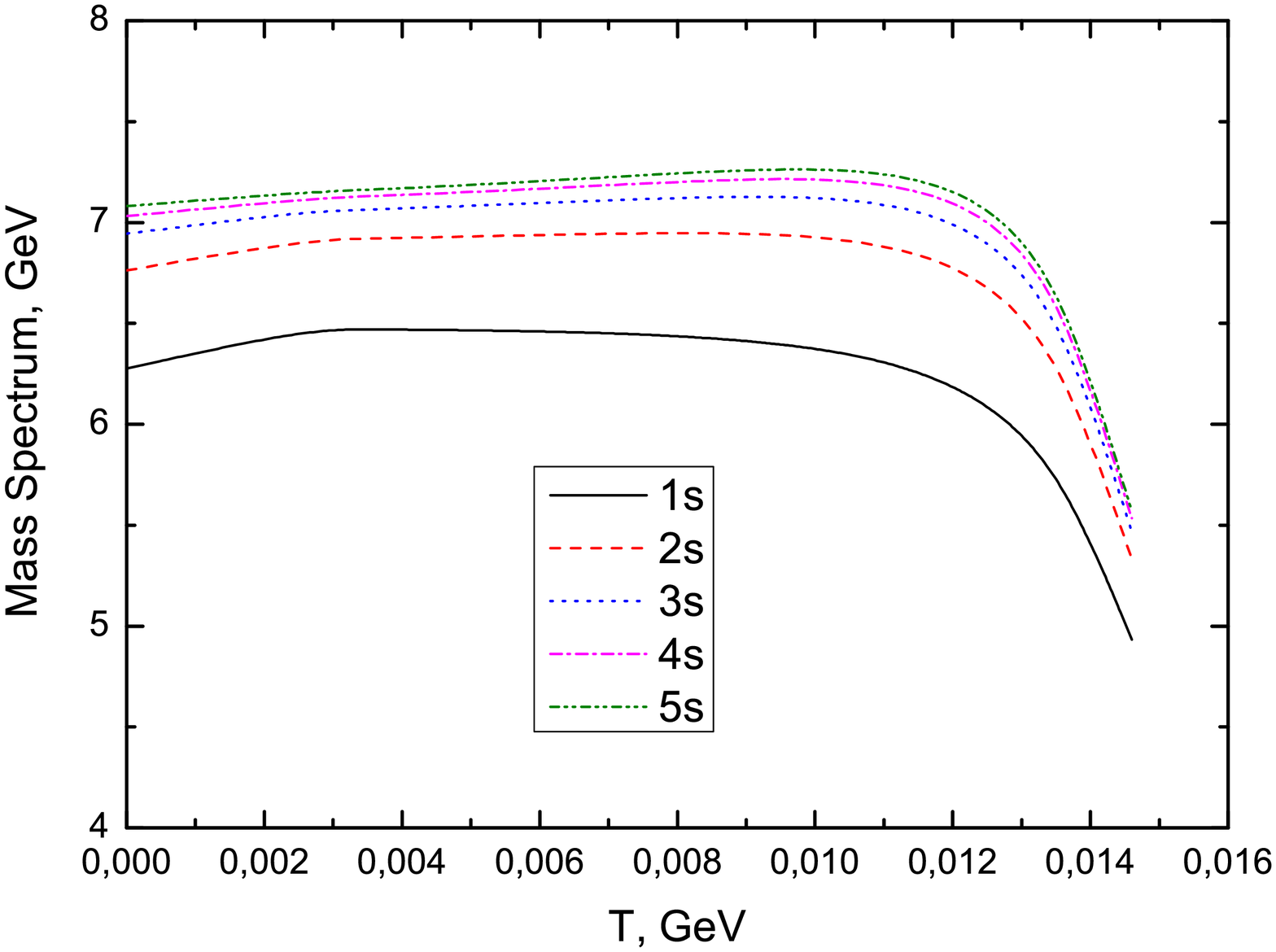}%\hspace{2pc}%
\caption{The mass spectrum of $b\overline{c}$  for $1s$, $2s$, $3s$,
$4s$, and $5s$ states as a function  of the temperature $T$ with
masses $m_c=1.209\,\, \text{GeV}$, $m_b=4.823\,\, \text{GeV}$, and
parameters   $A=0.147\,\, \text{GeV}^2$, $B=1.204$, $C=2.8\times
10^{-3}$, $D=2.4\times 10^{-5}$, and
$\delta=0.379\,\,\text{GeV}$.}\label{fig6}
\end{figure}

Firstly, we present the graphs for the change of the meson masses
within temperature dependent Cornell potential in Figs.\ref{fig1} -
\ref{fig3}. Afterwards, we present the same calculation for the
Cornell plus the inverse quadratic and harmonic potential in
Figs.\ref{fig4} - \ref{fig6}. In all these graphs, we see that
there exists a substantial decrement in the meson masses around
$T=120\,\text{MeV}$ which corresponds to $T=0.71\,T_c$. When we
increase the principal quantum number $n$, temperature leads an
increment in masses of the charmonium and $B_c$ mesons up to some
point and then there is a sharp decrement. However, this phenomena
is a bit different for the bottomonium, such that these states
firstly do not change and then start decreasing after some point.

Notice that although the number of the parameters for the Cornell
potential is less than the current potential, the results are not so
different. Thus, we may conclude that our observation for the
temperature-dependent masses does not depend much more on the number
of parameters. Besides that, our results for the temperature
dependent masses are in agreement with  QCD sum rule results for the
ground states \cite{Veliev}. One needs to perform more analyses for
other resonances in order to check the validity of this agreement
between pure nonrelativistic effects and QFT.

These results may  open  new possibilities for determining the
properties of the interactions in hadronic system. As a conclusion
of the results presented in these tables and figures  which are
based on analytically results, we may draw the following key
points: Firstly, temperature-dependent masses either for the Cornell
or the Cornell plus inverse quadratic and harmonic-type potential
are very sensitive to the choice of $n_r$ radial and $l$ orbital
quantum numbers. Secondly, we may consider the temperature-dependent
results valid because of the similar shapes between the Cornell and
current potentials. These results are sufficiently accurate for
practical purposes.

\section{Conclusion}\label{6}
The temperature-dependent  Schr\"{o}dinger equation is investigated
by applying the  NU method . As a potential part of the
Schr\"{o}dinger equation, the Cornell plus  inverse quadratic and
harmonic oscillator-type is used. Analytical expression for the
energy eigenvalues and the radial wave functions is presented.
Results are used for describing nonzero and zero temperature mass
spectra of heavy quarkonia and $B_{c}$. Numerical results are
compared  with the experimentally well-established resonances, and
some predictions are presented for the states which have not been
confirmed yet. For instance, we have the predictions for
bottomonium, namely, beside $\Upsilon(1s)$, $\Upsilon(2s)$,
$\Upsilon(3s)$, and $\Upsilon(4s)$ resonances, we predict the
resonance $\Upsilon(10860)$ and $\Upsilon(11020)$ to be $7s$ and
$10s$ states, respectively. For the charmonium states, besides the
$J/\psi(1s)$ and $\psi(2s)$ resonances, we predict $3s\longrightarrow
\psi(4040)$, $4s\longrightarrow \psi(4260)$, and $5s\longrightarrow
\psi(4415)$. We have seen that a zero temperature mass spectrum for
$l=0$ is quite in agreement with current experimental data, while
for $l\neq 0$, there is some disagreement with experimental data
because of the missing spin-spin and spin-orbital momentum
interactions terms within the potential. For the temperature-
dependent case, we see the strong dependence of the quarkonia mass
spectrum on the quantum numbers. Temperature-dependent results for
ground states are in good agreement with quantum field theoretical
approach such as QCD sum rules results. We have seen that  the
extension of the Cornell potential leads to slight changes on the
masses for nonzero temperature as well as zero temperature.

It seems that this simple potential model, like other nonrelativistic
models, is not enough to describe all  the features of hadrons
within a  thermal effect such that having all the hadrons melted at the
same temperature contradicts with the lattice data. The most
convenient way to compare the prediction of potential models with a
direct calculation of quarkonium spectral functions is to calculate
the Euclidean meson correlator at finite temperature to compare the
lattice data as it was done for the Cornell potential in
\cite{Mocsy:2004bv,Mocsy:2005qw}.  It has been shown that even
though potential models with certain screened potentials can
reproduce qualitative features of the lattice spectral function,
such as the survival of the $1s$ state and the melting of the $1p$
state, the temperature dependence of the meson correlators is not
reproduced. According to the lattice results
\cite{Umeda:2002vr,Asakawa:2003re} the $1s$ charmonium survives up
to $1.5 T_c$ and the $1p$ charmonium dissolves by $1.16 T_c$ and
higher excited states disappear near the transition temperature. It
is possible that the effects of the medium on quarkonia binding
cannot be understood in a simple potential model. However, one can
still do this comparison for other potential models such as the one
we have used in this paper. One of the main step for correlator
calculation is about the exact solution of the radial
Schr\"{o}dinger equation which is already done in this paper.

The method used in this paper are the systematic ones, and in many
cases, it is one of the most concrete works in this area. In
particular, the extended Cornell  potential can be one of the
important potentials, and it deserves special concern in many
branches of physics, especially in hadronic, nuclear and  atomic
physics.

Consequently, studying for an analytical solution of the modified
radial Schr\"{o}dinger equation for the   sum of the Cornell,
inverse quadratic, and harmonic-type potential within the framework of
ordinary quantum mechanics  could provide valuable information on
the elementary particle  physics and quantum chromodynamics  and
opens new windows for further investigation.

We can conclude that the theoretical  results of this study are
expected to enable new possibilities for pure theoretical and
experimental physicist, because of the exact and more general nature
of the results.

\section*{Data Availability}
The information given in our tables is available for readers in
the original references listed in our work.
\section*{Conflicts of Interest}
The authors declare that they have no conflicts of interest.
\section*{Funding}
We would like to note that, current research did not receive any specific funding but was performed as part of the employment of the all three authors.
\section*{Acknowledgement}

We would like to thank Shahriyar Jafarzade for careful reading the
manuscript.

%\newpage
%\clearpage


\begin{thebibliography}{99}

\bibitem{Appelquist75}
T. Appelquist and H. D. Politzer,  "Heavy quarks and $e^+e^-$
annihilation", Phys. Rev. Lett. \textbf{34}, pp.43-45,  1975.
\bibitem{Greiner}
 W. Greiner, Quantum Mechanics. Springer, Berlin, 2001.
\bibitem{Bagrov}
V. G. Bagrov, D. M. Gitman, Exact Solutions of Relativistic Wave
Equations (KluwerAcademic Publishers, Dordrecht, 1990)
\bibitem{Landau}
L. D. Landau and E. M. Lifshitz, Quantum Mechanics, NonRelativistic
Theory. 1977.
\bibitem{Dong}
Shi-Hai Dong, Factorization Method in Quantum Mechanics. Springer,
Dordrecht, 2007.
\bibitem{Kuchin}
S. M. Kuchin and N. V. Maksimenko,  "Theoretical estimations of the
spin averaged mass spectra of heavy quarkonia and bc mesons"  Univ.
J. Phys. Appl. \textbf{1(3)}: pp.295-298,  2013.
\bibitem{Al-Jamel}
A. F. Al-Jamel and H. Widyan,  "Heavy quarkonium mass spectra in a
Coulomb field plus quadratic potential using Nikiforov-Uvarov
method," Appl. Phys. Res. vol.\textbf{4}, no.3, pp.94-99, 2012.
\bibitem {Maksimenko}
N. Maksimenko and S. Kuchin, "Determination of the mass spectrum of
quarkonia by the Nikiforov-Uvarov method,"  Russ. Phys. J. \textbf{54}, no.1,
pp.57-65, 2011.
\bibitem{Ghalenovi}
Z. Ghalenovi, A. A. Rajabi, S. Qin, and D. H. Rischke,
"Ground-state masses and magnetic moments of heavy baryons," Mod.
Phys. Lett. A \textbf{29},  1450106, 2014.
\bibitem {Vigo-Aguir}
B. J. Vigo-Aguiar and T. E. Simos, " Review of multistep methods for
the numerical solution of the radial Schr\"odinger equation," Int.
J. Quantum Chem. vol. \textbf{103}, no.3, pp.278-290,  2005.
\bibitem{Ahmadov1}
M. Abu-Shady, H. Mansour, and A. I. Ahmadov, "Dissociation of
quarkonium in hot and dense media in an anisotropic plasma in the
nonrelativistic quark model," Adv. High Energy Phys. vol. \textbf{2019},
Article ID 4785615, 2019.
\bibitem{Karayer1}
H. Karayer,  "Study of the radial Schr\"odinger equation with
external magnetic and AB flux fields by the extended
Nikiforov-Uvarov method", Eur. Phys. J. Plus \textbf{135}, 70, 2020.
\bibitem {Karayer2}
H. Karayer, D. Demirhan and F. B{\"u}y{\"u}kk{\i}l{\i}{\c{c}}, Solution of
Schr\"odinger equation for two different potentials using extended
Nikiforov-Uvarov method and polynomial solutions of biconfluent Heun
equation," Journal of Mathematical Physics \textbf{59} no. 5, 053501, 2018.
\bibitem {Matsui}
T. Matsui and H. Satz, "$J/\psi$   Suppression by quark-gluon plasma
formation," Phys. Lett. B \textbf{178}, pp.416-422, 1986.
\bibitem {Karsch}
 F. Karsch, M. Mehr, and H. Satz, "Color screening and deconfinement for bound states of
heavy quarks," Z. Phys. C \textbf{37}, 617, 1988.
\bibitem {Aby-Shady}
M. Abu-Shady, "N-dimensional   Schr\"odinger equation at finite
temperature using the Nikiforov-Uvarov method," Journal of the
Egyptian Mathematical Society \textbf{25}, pp.86-89,  2017.
\bibitem {Fingberg}
 J. Fingberg, "Heavy quarkonia at high temperature," Phys. Lett. B \textbf{424}, pp. 343-354,  1998.
\bibitem {El-Naggar}
N. El-Naggar, L. Abou Salem, A. Shalaby, and M. Bourham, "The
equation of state for non-ideal quark gluon plasma," Phys. Sci. Int.
J. \textbf{4} no. 7, pp.912-929, 2014.
\bibitem {Malik}
G. P. Malik, R. K. Jha, and V. S. Varma,  "Finite-temperature
Schr\"odinger equation: solution in coordinate space," The Astrophysical
Journal, \textbf{503}, 446,  1998.
\bibitem {Wu}
X.-Y. Wu, B.-J. Zhang, X.-J. Liu, Y.-H. Wu, Q.-C. Wang, and
Y. Wang, "Finite temperature Schr\"odinger equation," Int. J.
Theor. Phys. \textbf{50}, pp.2546-2551, 2011.
\bibitem {Maireche}
A. Maireche, "A theoretical investigation of nonrelativistic bound
state solution at finite temperature using the sum of modified
Cornell plus inverse quadratic potential," Sri Lankan Journal of
Physics \textbf{21} no. 25, pp.11-36, 2020.
\bibitem {Ahmadov2}
A. Ahmadov, C. Aydin, and O. Uzun, "Bound state solution of the
Schr\"odinger  equation at finite temperature," J. Phys. Conf. Ser.
\textbf{1194}, no. 1,  012001, 2019.
\bibitem {Abu-Shady3}
 M. Abu-Shady and A. Ikot, "Analytic solution of multi-dimensional Schr\"odinger  equation in
hot and dense QCD media using the SUSYQM method," Eur. Phys. J. Plus
\textbf{134}, no. 7, 321, 2019.
\bibitem {Abu-Shady4}
M. Abu-Shady and A. Ikot, "Dissociation of nucleon and heavy-baryon
in an anisotropic hot and dense QCD medium using Nikiforov-Uvarov
method," Eur. Phys. J. Plus \textbf{135}, no. 6,  406, 2020.
\bibitem {Alberico:2006vw}
W. M. Alberico, A. Beraudo, A. De Pace, and A. Molinari,  Quarkonia
in the deconfined phase: effective potentials and lattice
correlators," Phys. Rev. D \textbf{75}, 074009, 2007.
\bibitem {Eichten1}
E. Eichten, K. Gottfried, T. Kinoshita, J. B. Kogut, K. Lane, and
T.-M. Yan, "Spectrum of charmed quark-antiquark bound states," Phys.
Rev. Lett. \textbf{34}, pp.369-372, 1975.
\bibitem {Eichten2}
E. Eichten, K. Gottfried, T. Kinoshita, K. Lane, and T.-M. Yan,
"Charmonium: the model," Phys. Rev. D \textbf{17}, no.11, 3090,
1978.
\bibitem {Rani}
 R. Rani, S. Bhardwaj, and F. Chand, "Mass spectra of heavy and light mesons using
asymptotic iteration method," Commun. Theor. Phys. \textbf{70}, no. 2,  179,
2018.
\bibitem {Abu-Shady1}
M. Abu-Shady, "Heavy quarkonia and $B_c$-mesons in the Cornell
potential with harmonic oscillator potential in the N-dimensional
Schr\"odinger equation," APS Physics \textbf{2}, pp.16-20,  2016.
\bibitem {Vega}
A. Vega and J. Flores, "Heavy quarkonium properties from Cornell
potential using variational method and supersymmetric quantum
mechanics", Pramana \textbf{87}, 73, 2016.
\bibitem {Khokha}
 E. Khokha, M. Abu-Shady, and T. Abdel-Karim, "Quarkonium masses in the N-dimensional
space using the analytical exact iteration method", International
Journal of Theoretical and Applied Mathematics, vol.\textbf{2}, no.2,pp.
86-92, 2016.
\bibitem {Ezz-Alarab}
M. Abu-Shady, T. Abdel-Karim, and S. Y. Ezz-Alarab, "Masses and
thermodynamics properties of heavy mesons in the non-relativistic
quark model using Nikiforov-Uvarov meethod", J. Egyptian Math. Soc.
\textbf{27}, 14,  2019.
\bibitem {Gamal}
H. Mansour and A. Gamal, Bound state of heavy quarks using a general
polynomial potential," Adv. High Energy Phys. vol. \textbf{2018}, Article ID
7269657, 2018.
\bibitem {Abu-Shady2}
M. Abu-Shady, T. Abdel-Karim, and E. Khokha, "Exact solution of the
$N$-dimensional radial Schr\"odinger equation via Laplace
transformation method with the generalized Cornell potential",
SciFed Journal of Quantum Physics, vol.\textbf{2}, no.1, pp.1-11, 2018.
\bibitem {Hassanabadi}
S. Hassanabadi, A. Rajabi, and S. Zarrinkamar,  "Cornell and Kratzer
potentials within the semirelativistic treatment", Mod. Phys. Lett.
A \textbf{27}, 1250057, 2012.
\bibitem {Ikot}
 A. N. Ikot, G. J. Rampho, P. O. Amadi, U. S. Okorie, M. J. Sithole, and M. L. Lekala, "Quantum
information-entropic measures for exponential-type potential,"
Results in Physics, \textbf{18}, 103150,  2020.
\bibitem {Dong3}
Shi-Hai Dong, Guo-Hua Sun, and M. Lozada-Cassou, "Exact solutions
and ladder operators for a new anharmonic oscillator," Physics
Letters A. \textbf{340}, 94,  2005.
\bibitem {Chen3}
 Xiao-Hua Wang, Shi-Hai Dong, Chang-Yuan Chen, and Yuan You,  "Exact solutions of the
Schr\"odinger equation with double ring-shaped oscillator," Physics
Letters A. \textbf{377}, 1521, 2013.
\bibitem {Dong4}
 Qian Dong, H.Ivn\'an Garcia Hern\'andez, Guo-Hua Sun, Mohamad Toutounji, and
Shi-Hai Dong,   "Exact solutions of the harmonic oscillator plus
non-polynomial interaction," Proc. R. Soc. A \textbf{476},  20200050, 2020.
\bibitem {Dong5}
Xi-Wen Hou, Shi-Hai Dong, and Zhong-Qi Ma,  "Relativistic levinson
theorem in two Dimensions," Phys. Rev. A. \textbf{58}, no. 3,  2160, 1998.
\bibitem {Nikiforov}
 A. F. Nikiforov and V. B. Uvarov, Special Functions of Mathematical Physics, Birkh\"{a}user, Basel 1988.
\bibitem {Pekeris}
C. Pekeris, "The rotation-vibration coupling in diatomic molecules,"
Phys. Rev. \textbf{45}, pp.98-103, 1934.
\bibitem {Omugbe}
E. Omugbe, O. E. Osafile, and M. C. Onyeaju,  Mass spectrum of
mesons via the WKB approximation method," Adv. High Energy Phys.,
vol.\textbf{2020}, Article ID  5901464, 2020.
\bibitem {Zyla}
P.A. Zyla et al. (Particle Data Group), Prog. Theor. Exp. Phys.
\textbf{2020}, 083C01 (2020).
\bibitem {Faustov}
R. N. Faustov, V. O. Galkin, A. V. Tatarintsev, and A. S. Vshivtsev, "Spectral
problem of the radial Schr\"odinger equation with confining power
potentials," Theor. Math. Phys. \textbf{113}, pp.1530-1542,  1997.
\bibitem{Kumar}
R. Kumar and F. Chand, "Series solutions to the N-dimensional radial
Schr\"odinger equation for the quark-antiquark interaction
potential,"   Phys. Scr. \textbf{85}, 055008,  2012.
\bibitem {Boyd}
G. Boyd, J. Engels, F. Karsch, E. Laermann, C. Legeland, M.
Lutgemeier, and B. Petersson, "Thermodynamics of SU(3) lattice gauge
theory," Nucl. Phys. B \textbf{469},  pp.419-444, 1996.
\bibitem {Bernard}
C. Bernard, T. Burch, E. Gregory  et al., "QCD
thermodynamics with three flavors of improved staggered quarks,"
Phys. Rev. D \textbf{71}, 034504, 2005.
\bibitem {Veliev}
E. Veliev, H. Sundu, K. Azizi, and M. Bayar,  "Scalar quarkonia at
finite temperature," Phys. Rev. D \textbf{82}, 056012, 2010.
\bibitem {Mocsy:2004bv}
\'A. M\'ocsy and P. Petreczky, "Heavy quarkonia survival in
potential model,"  Eur. Phys. J. C \textbf{43}, pp. 77-80, 2005.
\bibitem {Mocsy:2005qw}
 \'A. M\'ocsy and P. Petreczky, "Quarkonia correlators above deconfinement," Phys. Rev. D \textbf{73}, 074007, 2006.
\bibitem {Umeda:2002vr}
T. Umeda, K. Nomura, and H. Matsufuru,  "Charmonium at  finite
temperature in quenched lattice QCD," Eur. Phys. J. C \textbf{39S1}, pp.
9-26, 2005.
\bibitem {Asakawa:2003re}
M. Asakawa and T. Hatsuda, $J/\psi$ and $\eta_c$ in the deconfined
plasma from lattice QCD,  Phys. Rev. Lett. \textbf{92}, 012001, 2004.

\end{thebibliography}
\end{document}